\documentclass[aps,pre,reprint,superscriptaddress,amsmath]{revtex4-1}
\usepackage{breqn}
\usepackage{array}
\usepackage{units}
\usepackage{graphicx}
\usepackage{hyperref}
\DeclareMathOperator\arctanh{arctanh}
\begin{document}

%\preprint{}

\title{Effects of cell elasticity on the migration behavior of a monolayer of motile cells: Sharp Interface Model}


\author{Yony Bresler}
\author{Benoit Palmieri}
\author{Martin Grant}
%\affiliation{Department of Physics McGill University}
\affiliation{Department of Physics McGill University 3600 University Montr\'eal Qu\'ebec Canada H3A 2T8}

%Department of Physics, McGill University, 3600 University, Montr\'eal, Qu\'ebec, Canada H3A 2T8.

%\thanks{}
\email{martin.grant@mcgill.ca}


\date{\today}

\begin{abstract}
In order to study the effect of cell elastic properties on the behavior of assemblies of motile cells, this paper describes an alternative to the cell phase field (CPF) \cite{Palmieri2015} we have previously proposed. The CPF is a multi-scale approach to simulating many cells which tracked individual cells and allowed for large deformations. Though results were largely in agreement with experiment that focus on the migration of a soft cancer cell in a confluent layer of normal cells \cite{Lee2012}, simulations required large computing resources, making more detailed study unfeasible. In this work we derive a sharp interface limit of CPF, including all interactions and parameters. This new model offers over $200$ fold speedup when compared to our original CPF implementation. We demonstrate that this model captures similar behavior and allows us to obtain new results that were previously intractable. We obtain the full velocity distribution for a large range of degrees of confluence, $\rho$, and show regimes where its tail is heavier and lighter than a normal distribution. Furthermore, we fully characterize the velocity distribution with a single parameter, and its dependence on $\rho$ is fully determined. Finally, cell motility is shown to linearly decrease with increasing $\rho$, consistent with previous theoretical results.
\end{abstract}

%\keywords{}

\maketitle

%\section{}
%\subsection{}
%\subsubsection{}

\section{Introduction\label{sec:intro}}

Cell monolayers have been used to study a variety of biological process,
such as cancer metastasis, wound healing, colony fronts, immunosurveillance and collective
cell migration \cite{Trepat2009, Tambe2011, Friedl2008}. These can exhibit complex behavior which can be studied
at the intersection of in vivo, in vitro and in silico experiments.
This paper is motivated by the metastasis pathway where
a single cancer cell, having left the primary tumor, squeezes through
the much stiffer endothelium in an attempt to reach a nearby blood
vessel.

One method to simulate such cells which has been introduced in recent
years is a phase-field approach \cite{Nonomura2012,Najem2016,Camley2014,Lober2015},
where each cell, labeled $n$, is described by a field $\phi_{n}\text{\ensuremath{\left(\mathbf{r,t}\right)}}$.
These fields are defined at every point in space $\mathbf{r}$ and
time\textbf{ $\mathbf{t}$}, and smoothly transitions from unity
inside the cell to zero outside. In particular, the authors of this
paper previously developed a multi-scale phase field for studying
elasticity mismatch in cells \cite{Palmieri2015}.
That approach explicitly modeled each cell with tunable elasticity
and allowed for large deformations. It was argued that these deformations
play a key role in velocity ``bursts'', as a highly deformed cell
propagates to a more relaxed state with high velocity. However, a
major limitation of the model is the large computational resources
required for large-scale, long-time simulations that are needed to
obtain sufficient statistics.

There are several methods known for improving the efficiency of phase
field models. One approach is to replace the uniform mesh with an
adaptive mesh \cite{Provatas2013}. While this reduces the number
of mesh points required to perform a simulation, these methods still
explicitly track details in both the interior and exterior the cell. Here
we are only interested in the motion of the interface and simulating
the bulk can be superfluous. Additionally, when the cell is highly
deformed, the advantage of such an approach diminishes. Sometimes, in simple cases, it 
is possible to consider only the motion of the interface, and disregard the bulk evolution. 
See, for example \cite{Yao1992}, for a treatment of Ostwald ripening.
Another approach is to represent all cells with a single field using a vacancy phase
field crystal approach \cite{Alaimo2016}. This allows efficient modeling
of many cells but does not allow for large deformations.

A multitude of other models that have been used to study
the motion of cells. These include the Cellular Potts model approaches
to the motion of cells, including their interaction with the extracellular
matrix \cite{Galiano2010}; self-propelled Voronoi models to study
jamming transitions \cite{Bi2015,Bi2016} and swarm migration \cite{Grossman2008};
sub-cellular description using beads and springs models for cell rheology
\cite{Mkrtchyan2014} and tissue growth \cite{Mkrtchyan2014}; and continuum models 
to model unconstrained spreading of epithelial layers \cite{Kopf2013} or wound closure \cite{Lee2011}.
See \cite{Camley2017,Hakim2017} for recent comprehensive reviews of various methods
in collective cell motility. All these methods differ in their level
of description, but none both explicitly track each cell as well as
allow for tunable elasticity and arbitrarily large deformations. Furthermore, many of these
methods are limited to the study of perfectly confluent layers, whereas
we wish to address any degree of confluence. One notable exception is recent treatment 
by Madhikar \textit{et al. }\cite{Madhikar2018} which we became aware of after completing this work.

The model proposed in this paper bridges a gap between previous approaches.
As a sharp-interface limit of the CPF, there is a significant numerical
speedup over the standard CPF approach. At the same time, it retains
the description of single cells and allows for large deformations
with tunable elasticity. Since the 1950s, there has been extensive
work on sharp interface models, including approaches to the Stefan
problem \cite{rubinshtein1971stefan}, Ostwald ripening \cite{Voorhees1985}, and
the Gibbs-Thomson effect \cite{Johnson1965}. The phase field model
itself emerged to solve some of the morphological instabilities of
sharp interfaces. A review of this evolution can be found here \cite{Sekerka2004}.
The model we propose does not suffer from those instabilities, as
the number of cells is fixed, and cell volume is kept constant to
good approximation. In the absence of growth, the sharp interface
model is tractable and allows for a significant speedup over phase
field models.

There has also been previous work using a sharp-interface approach
to model or detect cell motion. Lee \cite{Lee1993a} used a sharp-interface
mathematical model to track the motion of single Keratocytes cells.
Their work has been extended to include details of the actin including
nucleation \cite{Grimm2003} and cell polarization \cite{Kozlov2007},
and the Filament Based Lamellipodium Model \cite{Manhart2015}. These
approaches provide a more detailed description of cells than the model
we propose. By their nature, these models are more specialized to
certain types of cells and may not be applicable to different cell
lines. They have also been used to model single cells, but not multi-cell
collective behavior.

The remainder of this paper is structured as follows: First we derive
our sharp interface model from the CPF including non-local terms,
model parameters, as well as outline implementation details. Section
\ref{Simulation-Results} is the results section, which includes reproducing 'bursts' seen
in the CPF, as well as a study of the effects of different concentrations
and elasticities on cell motion which were not feasible with CPF. We
conclude by summarizing our results and suggest future work.

\section{Sharp Interface Model for Cells}

\label{sharp-interface-model-for-cells}

We begin with the CPF model, where the time evolution for the field
of each cell (labeled n), $\phi_{n}$ is given in dimensionless units
by 

\begin{dmath}
\frac{d\phi_{n}}{dt}  =  -\textbf{v}_{n}\cdot\mathbf{\nabla}\phi_{n}+\gamma\mathbf{\nabla}^{2}\phi_{n}
 -\frac{30}{\lambda^{2}}\left[\gamma\phi_{n}\left(1-\phi_{n}\right)\left(1-2\phi_{n}\right)+2\kappa\sum_{m\neq n}\phi_{n}\phi_{m}^{2}\right]
  -\frac{2\mu}{\pi R_{0}^{2}}\phi_{n}\left[\int dx\int dy\:\phi_{n}^{2}-\pi R_{0}^{2}\right],\label{eq:phasefieldcells}
\end{dmath}
where $\textbf{v}_{n}$ is the velocity of each cell, $\gamma$ is
the parameter that controls cell stiffness or elastic response, $\lambda$ is the width
of the cell boundary interface, $\kappa$ sets the strength of neighbor-neighbor
cell repulsion, and $\mu$ is a soft constraint to keep cells at their
preferred size, $\pi R_{0}^{2}$. The interior of a cell is described
where $\phi=1$, smoothly decreasing across the interface, until reaching
$\phi=0$ at the cell exterior. This phase field model was shown to
be successful, though it is computationally taxing. Hence, we set out
to take the sharp interface limit of the phase field equations. 

Consider a phase field with a free-energy function given by 
\begin{equation}
F=\int d^{d}r\left[\frac{1}{2}C\left(\nabla\phi\right)^{2}+f\left(\phi\right)\right],\label{eq:phase-field-generic}
\end{equation}
where C is a constant, and $f\left(\phi\right)$ is the bulk free
energy which has any double well structure. When the thickness of
the interface is much less than the radius of curvature of the interface,
the local interface velocity $\mathbf{v}_{interface}$ can be approximated \cite{Grant1983}
as 
\[
\mathbf{v}_{interface}=\Gamma K\hat{\mathbf{n}}.
\]
Here, $K$ denotes the local curvature, $\Gamma$ is a model dependent
parameter, and $\mathbf{v}_{interface}$ points along the local normal to the interface,
denoted as $\hat{\mathbf{n}}$. Note that the interface velocity is
entirely independent of the choice of $f(\phi)$. Now consider a system
governed by a free-energy of the form shown in Equation \ref{eq:phase-field-generic}
and an interface initially given by $R\left(\theta,t=0\right)$, where
$R\left(\theta,t\right)$ describes the distance to the interface
from the cell center for any angle $\theta$. We shall use this angular 
representation for simplicity, and though it precludes multi-valued radii, 
we will later generalize to remove this restriction. Its time evolution
will be given by 
\begin{equation}
\left.\frac{\partial\mathbf{R}\left(\theta,t\right)}{\partial t}\right|_{Curvature}=\mathbf{v}_{interface}(\theta,t)=\gamma K\left(\theta,t\right)\mathbf{\hat{n}}\left(\theta,t\right),\label{eq:drdt curvature}
\end{equation}
where in this model, $\Gamma=\gamma$. Evolving the system in time
is thus reduced to computing the local curvature along the interface.

\subsection{Approximating non-local terms}

Although the sharp interface limit of Equation \ref{eq:phase-field-generic}
does not depend on the details of the bulk free energy $f(\phi)$,
the same does not hold for non-local terms. The CPF model had two
such terms, the area conservation, and neighbor-neighbor interaction
terms. Similar to the sharp interface approximation, we assume that
$\lambda \ll K$ such that the radial
component is in 1D equilibrium and obtain sharp-interface estimate
of these terms by calculating the rate of change of the location of
the interface. The long-time equilibrium solution of Equation \ref{eq:phasefieldcells}
for a cell interface centered at $x=0$ is given by 
\begin{equation}
\phi_{n}^{*}(x)=\frac{1+\tanh\left(\alpha x\right)}{2},\label{eq:PFequilibriumSoln}
\end{equation}
with $\alpha=\sqrt{\frac{15}{2}}/\lambda$. We define the sharp interface
to be at the point $x=0$, where $\phi_{n}^{*}(x)=0.5$ is halfway
between the interior ($\phi=1$) and exterior ($\phi=0$) the cell.

Recall the evolution of the phase field model is restricted by the
area conservation term 
\[
\left.\frac{d\phi_{n}}{dt}\right|_{Area}=-\frac{2\mu}{\pi R^{2}}\phi_{n}\left[\int dx\int dy\phi_{n}^{2}-\pi R_{0}^{2}\right].
\]
Consider the interface $\phi_{n}$ to be at equilibrium at time $t_{i}$
but with a change $\Delta A$ from the equilibrium area $A_{eq}=\pi R_{0}^{2}$.
Assuming all other terms remain at equilibrium, this will introduce
a change to the field 
\[
\left.\frac{d\phi_{n}}{dt}\right|_{Area}=-\frac{2\mu}{\pi R^{2}}\phi_{n}^{*}\left[A_{eq}+\Delta A-A_{eq}\right]=-\frac{2\mu}{\pi R^{2}}\phi_{n}^{*}\Delta A.
\]
Using the 1st order (forward-Euler) approximation, we find that the
solution at $t_{i+1}$ is
\[
\phi_n(x,t_{i+1})=\phi_n^{*}(x)+\Delta t\cdot\left(-\beta\phi_{n}^{*}(x)\right),
\]
with $\beta=\frac{2\mu\Delta A}{\pi R^{2}}$. Solving for the updated
position of the interface, $\phi(x,t_{i+1})=0.5$, and using Equation
\ref{eq:PFequilibriumSoln} , the updated position of the interface
will be
\begin{dgroup*}
\begin{dmath*}
0.5 =\phi_n^{*}(x)+\Delta t\cdot\left(-\beta\phi_{n}^{*}(x)\right)
\end{dmath*}
\begin{dmath*}
x = -\frac{1}{\alpha}\arctanh\text{\ensuremath{\left(\frac{\beta\Delta t}{\beta\Delta t-1}\right)}}.
\end{dmath*}
\end{dgroup*}
 We then take the limit 
\[
\lim_{\Delta t\rightarrow0}\frac{x}{\Delta t}=\frac{\beta}{\alpha}=\sqrt{\left(8/15\right)}\lambda,
\]
 to obtain the rate at which a change in area $\Delta A$ moves the
position of the interface. Thus, we obtain an approximation for the
area conservation term in the sharp interface limit, 
\begin{equation}
\left.\frac{\partial\mathbf{R}\left(\theta,t\right)}{\partial t}\right|_{Area}=\mu'\left(A\left(\mathbf{R}\left(\theta,t\right)\right)-\pi R_{0}^{2}\right)\mathbf{\hat{n}}\left(\theta,t\right),\label{eq:drdt area}
\end{equation}
where we have defined $\mu'=\frac{\sqrt{\nicefrac{8}{15}}\lambda}{\pi R_{0}^{2}}\mu$.
Note that this derivation also assumed that the area of a cell at
equilibrium is $\pi R_{0}^{2}$, for both the sharp interface and
CPF models. This is a good approximation since $\int_{-\infty}^{d}((1+tanh(x))/2)^{2}\,dx\simeq d-0.5$.

To derive the neighbor-neighbor interactions, we follow a similar
procedure, though the equations do not have a closed form solution
and will require some additional approximations. We first consider
the interface of cell $n$ centered at $x=0$, and another cell with
an interface in the opposite orientation centered at $x=-d$. We assume
the cells to be at their unperturbed equilibrium, and now overlap.
In actuality, this overlap would perturb the interface resulting in
a smaller interface width and profile, though those corrections are
unnecessary for our desired level of description. Assuming all other
terms remain at equilibrium, this overlap leads to an updated interface
location

\begin{dgroup*}
\begin{dmath*}
\phi_{n}(x,t_{i+1})  =\phi_{n}^{*}(x)+\Delta t\frac{-30}{\lambda^{2}}2\kappa\phi_{n}^{*}(x){\phi_{m}^{*}(-x-d)}^{2}
\end{dmath*}
\begin{dmath*}
0.5  =\phi_{n}^{*}(x)\left[1-\Delta t\frac{30}{\lambda^{2}}2\kappa{\phi_{m}^{*}(-x-d)}^{2}\right]
\end{dmath*}
\begin{dmath*}
0.5  =\frac{\kappa}{2\lambda^{2}}\left\{ -15\Delta t+\lambda^{2}+30\Delta t\tanh(\alpha d)-15\Delta t\tanh^{2}(\alpha d)+ \left[15\alpha\Delta t+\alpha\lambda^{2}-45\alpha\Delta t\tanh^{2}(\alpha d)+30\alpha\Delta t\tanh^{3}(\alpha d)\right]x\right\} 
\end{dmath*}
\begin{dmath*}
x  =\frac{15\left(\Delta t-2\Delta t\tanh(\alpha d)+\Delta t\tanh^{2}(\alpha d)\right)}{\alpha\left(15\Delta t+\lambda^{2}/\kappa-45\Delta t\tanh^{2}(\alpha d)+30\Delta t\tanh^{3}(\alpha d)\right)},
\end{dmath*}
\end{dgroup*}

where in the 3rd line we have taken the 1st order series expansion
around $x=0$, so that the equation may be solved analytically. Taking
the same limit as in the area term gives us an expression for the
rate of change of the interface position due to the neighbor-neighbor
term
\[
\lim_{\Delta t\rightarrow0}\frac{x}{\Delta t}=\frac{15\kappa\left(\tanh(\alpha d)-1\right)^{2}}{\alpha\lambda^{2}},
\]
as a function of the distance $d$ between the two cell interfaces. Thus, the
sharp interface limit of the interactions between cell n and all other
cells is given by
\begin{align}
\left.\frac{\partial\mathbf{R}_{n}\left(\theta,t\right)}{\partial t}\right|_{Neigh} & =\frac{15\kappa\left(\tanh\left(\alpha d_{n,\theta'}\right)-1\right)^{2}}{\alpha\lambda}\hat{\mathbf{n}}\left(\theta,t\right),\label{eq:drdt neighbor}
\end{align}
where $d_{n,\theta'}$ is understood to be the distance between $\mathbf{R}_{n}\left(\theta,t\right)$
along $\mathbf{n}_{\theta}$, to the nearest neighboring cell.

\begin{figure}[tb]
\begin{centering}
\includegraphics[width=0.5\textwidth]{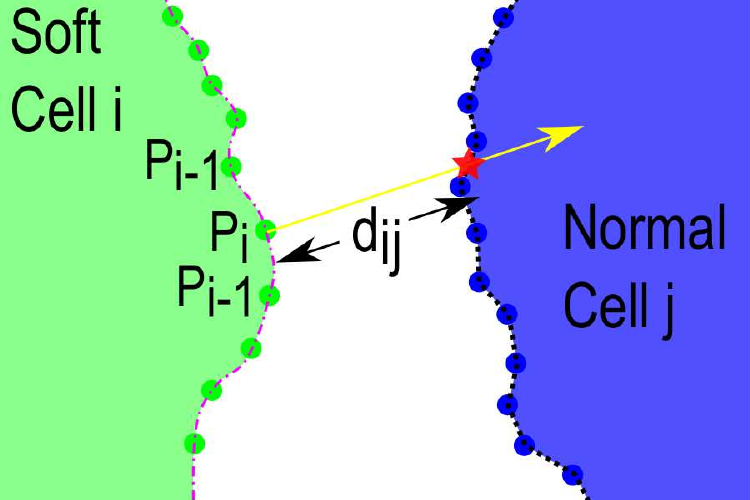}
\par\end{centering}
\caption{\textbf{Schematic of sharp interface model. }Two adjacent cells, a
soft (green, cell 'i') and normal (blue, cell 'j'). $P_{i}$'s are
the discrete positions of the interface. The local normal (yellow
line) is used to find the point of intersection (red star) with straight
segments adjoining positions along cell j (dashed black line). If
cell 'i' needed to be redistributed, it is evenly distributed along
the spline fit (pink dashed line). Features in this figure have been
exaggerated.}
\end{figure}

\subsection{Cell Velocity}

Up to this point, our description of the sharp interface has not accounted
for cell motility. We have already seen that a sharp interface cell
$n$ will move with a velocity $\mathbf{v}_{n}$ with the additional
term 
\[
\frac{\partial\mathbf{R}_{n}\left(\theta,t\right)}{\partial t}=\mathbf{v}_{n}.
\]
As in the CPF, each cell will have a single velocity which consists
of both an active, and an inactive part, 
\begin{equation}
\mathbf{v}_{n}=\mathbf{v}_{n}^{Active}+\mathbf{v}_{n}^{Inactive}.\label{eq:drdt v total}
\end{equation}
The active velocity describes the cell's net self propulsion, and
is driven by the cellular motors. This velocity can be applied directly
to each cell. As before, we chose this active velocity to have a constant
magnitude $v_{A}$ and reoriented with probability $P(t)=\frac{1}{\tau}e^{-t/\tau}$,
where $\tau$ is the mean time between reorientations. The inactive
velocity of cell $n$ is due to forces exerted by the other cells
surrounding it. In CPF, the inactive velocity was given by 
\[
\textbf{v}_{n}^{Inactive}=\frac{60\kappa}{\xi\lambda^{2}}\int dx\int dy\phi_{n}\left(\mathbf{\nabla}\phi_{n}\right)\sum_{m\neq n}\phi_{m}^{2},
\]
where $\xi$ is due to friction between the cells, the substrate,
and the surrounding water. In a similar procedure to estimating the
non-local terms, we will substitute the equilibrium solution $\phi^{*}(x)$
for both cells $n$ and $m$, and evaluate the expression as a function
of the distance $d$ between the two interfaces. No approximations or simulated time step is
needed, and the integral can be evaluated exactly to give
\begin{dmath}
\mathbf{v}_{n}^{Inactive}=\sum_{\theta'}\frac{30k}{\left(e^{2\alpha d_{n,\theta'}}-1\right)^{4}\lambda^{2}}\left[1+e^{4\alpha d_{n,\theta'}}\left(4\alpha d_{n,\theta'}-5\right)+e^{2\alpha d_{n,\theta'}}\left(4+8\alpha d_{n,\theta'}\right)\right]\mathbf{n}_{\theta'}.\label{eq:drdt v inactive}
\end{dmath}
Since the calculation of $d$ is the most computationally demanding
portion of our model, we use the distance $d_{n,\theta'}$ computed
for the neighbor-neighbor interaction, and assume that this value
of $\textbf{v}_{n}^{Inactive}$ is constant over the interval $\frac{\theta_{i}-\theta_{i-1}}{2}-\frac{\theta_{i+1}-\theta_{i}}{2}$.
Thus combining equations \ref{eq:drdt curvature}, \ref{eq:drdt area},
\ref{eq:drdt neighbor}, \ref{eq:drdt v total} \& \ref{eq:drdt v inactive}
the complete evolution of each cell is given by

\begin{dmath}
\frac{\partial\mathbf{R}_{n}\left(\theta,t\right)}{\partial t}=\left[K\left(\theta,t\right)+\mu'\left(A-\pi R^{2}\right)+\frac{150\left(\tanh\left(\alpha d_{n,\theta}\right)-1\right)^{2}}{\alpha\lambda}\right]\mathbf{n}_{\theta}+\mathbf{v}_{n}^{Inactive}+\mathbf{v}_{n}^{Active}.\label{eq:drdt total}
\end{dmath}

\subsection{Numerical Implementation}

To this point we have implicitly defined positions along the interface
by a radial representation, $R(\theta,t)$. While this is well suited
to cells that are mostly spherical, this representation has several
limitations, particularly for cells with large deformations. This representation
is explicitly single valued and cannot account for overhangs. There
are also challenges in the numerical implementation to the radial representation.
For example, to have uniform spacing in $\theta$, would require either
that all tangential components $\hat{\theta}$ be discarded to preserve
the uniform spacing; or alternatively that the points be redistributed
to be uniform after every time step.

For our model system and parameter set, we found that a Cartesian coordinate representation was more convenient. Hence, we moved to represent each
point $P_{i}$ along the interface by its position in the $x-y$ plane. This
representation requires
only infrequent redistribution of points. The curvature is computed as 
\[
K(P_{i})=-2\gamma\frac{P_{x}'P_{y}''-P_{y}''P_{x}''}{\left(\left(P_{x}'\right)^{2}+\left(P_{y}'\right)^{2}\right)},
\]
and the area is given by $A=\frac{1}{2}\int P_{x}P'_{y}\:dP$, where
$P_{x}'$ ($P_{y}'$ ) denotes a partial derivative along $x\,\left(y\right)$.
Since each point along the interface can move in any direction, they may move 
too closely together leading to numerical instability. Hence, we
redistribute the points along the interface by use of spline interpolation
when needed, using the centripetal Catmull-Rom spline \cite{Barry1988},
since it will not form closed loops or cusps within a curved section.
We perform this redistribution whenever adjacent points $\Delta P_{i}$
are either too close $(\Delta P_{i}<0.8\Delta\bar{P})$ or too far apart
$(\Delta P_{i}>1.5\Delta\bar{P})$, as compared to $\Delta\bar{P=}2\pi R/N$
which is the uniform spacing for a circle. 

Another challenge is that cells can become too deformed and lead to instabilities.
In particular, a soft cell that is pushed by two normal cells on opposites
sides, may 'pinch-off' a portion of the cell by bringing the two opposite
interfaces of the soft cell to touch. This can be resolved in several ways. The first is to make the cells stiffer, however
in our model system this would not allow for sufficient elastic mismatch
between the soft and normal cells. A second option is to penalize
pinching-off by adding a self repulsion term to the model. This would
better represent the mechanical and chemical mechanisms that exist
in real cells which prevent pinching-off from occurring. We opted
for a third approach, which is to offset the curvature term 
\begin{equation}
K_{natural}(P_{i})=K(P_{i})-\frac{1}{R_{0}}, \label{eq:Knatural}
\end{equation}
by the natural curvature of a cell with radius $R_{0}$, analogous to Helfrich theory \cite{Helfrich1973}. We found this was sufficient
to prevent the cell from being pinched-off while still allowing large deformations
needed for bursts. 

\begin{figure}[tb]
\includegraphics[width=0.9\columnwidth]{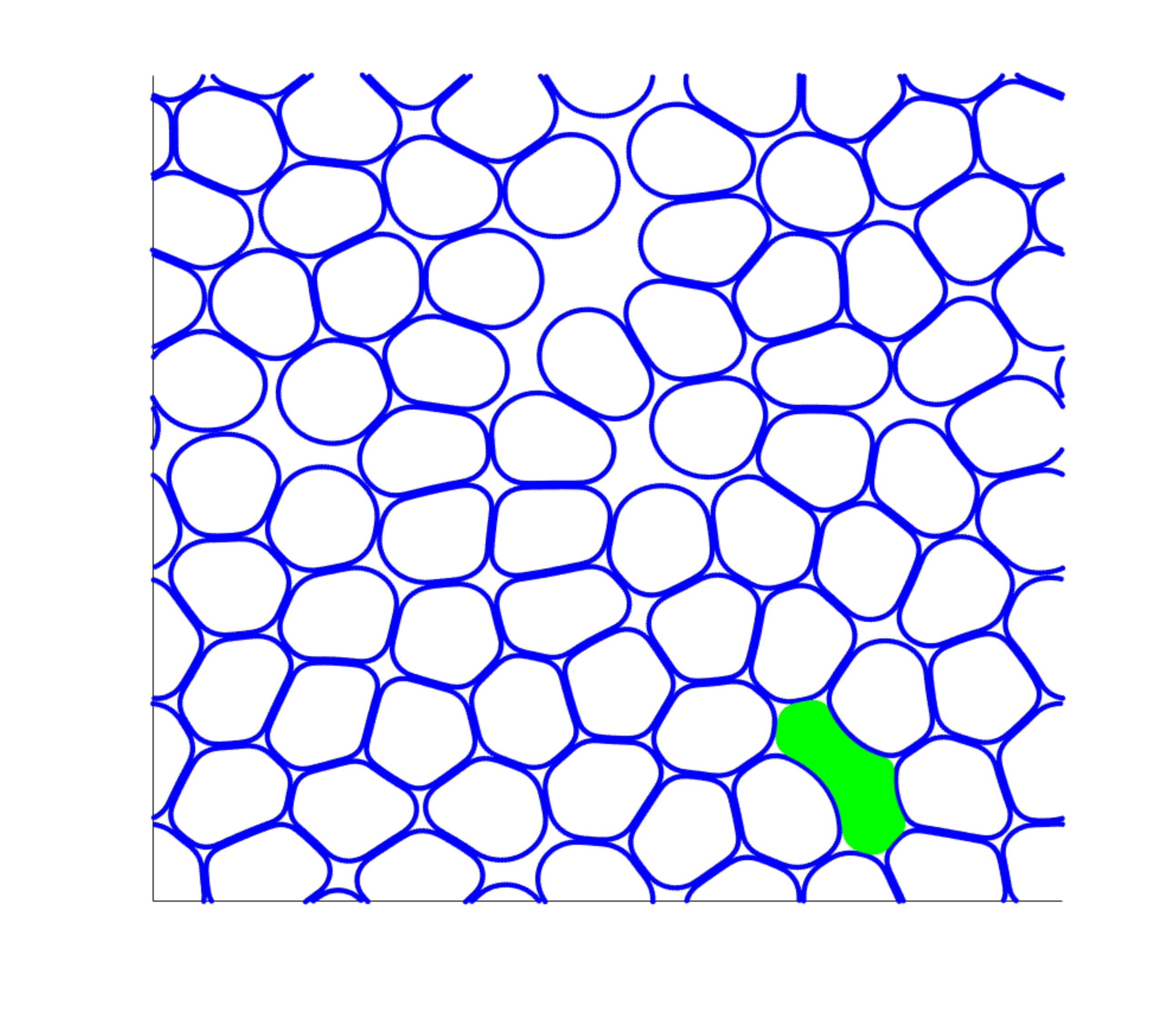}

\caption{\textbf{Simulation Snapshot. }Snapshot from a simulation with degree of confluence $\rho=80\%$,
showing the soft cell in green, and normal cells interfaces in blue.
\label{fig:SimulationSnapshot}}
\end{figure}

As shown in Table \ref{tab:Simulations-Parameters.-Table} model parameters
were similar to CPF, with some minor modifications, improving
the stability of cells while also preserving the deformation needed
to allow for bursts. Despite these minor changes, having derived our model directly from CPF, we think these models yields asymptotically the same results. A sample snapshot from a simulation is shown
in Figure \ref{fig:SimulationSnapshot}, where the soft cell is colored
in green, and all other cells have normal stiffness and are shown in
blue. 

% \begin{table}%[H] add [H] placement to break table across pages
% \begin{ruledtabular}
% Lines of table here ending with \\
% \end{tabular}
% \end{ruledtabular}
% \end{table}

\begin{table}[tb]
%\begin{ruledtabuler}
\noindent\resizebox{\columnwidth}{!}{%%
\begin{tabular}{|>{\centering}p{0.15\columnwidth}|>{\centering}p{0.22\columnwidth}|>{\centering}p{0.05\columnwidth}|>{\centering}p{0.05\columnwidth}|>{\centering}p{0.1\columnwidth}|>{\centering}p{0.05\columnwidth}|>{\centering}p{0.1\columnwidth}|}
\hline 
 & $\gamma$ & $\kappa$ & $\mu$ & $\xi$ & $\tau$ & $v_{A}$\tabularnewline
\hline 
CPF & soft: 0.3 \\
normal: 1 & 10 & 1 & $1.5\cdot10^{3}$ & $10^{4}$ & $10^{-2}$\tabularnewline
\hline 
Sharp Interface Model & soft: 0.45 \\
normal: 1.25  & 5 & 0.5 & $10^{3}$ & $10^{4}$ & $10^{-2}$\tabularnewline
\hline 
\end{tabular}}
\caption{\textbf{Model Parameters. }Table summarizing simulation parameters
for our sharp interface model, and comparison with CPF.\label{tab:Simulations-Parameters.-Table}}
%\end{ruledtabular}
\end{table}

Simulations were performed with the same parameters shown above, unless
otherwise mentioned, in a simulation box with periodic boundary conditions.
$72$ cells were used, each consisted of $N=150$ points, and forward Euler integration was used with a
time step $dt=0.1$. The curvature was computed
using a 2nd order (5 point) symmetric stencil. Simulations were initialized with the 
same random number seed and in a hexagonal lattice, and equilibration consisted of the first $t=40,000$. Following that, samples were taken every $t=800$, for a total simulation time of $t=2\cdot10^{6}$ for each run. Using these units, $t=1$
corresponds to roughly $0.36$s in real time. Our sharp interface
model was implemented using C++ and openMP, with use of the boost
library \cite{schaling2011boost}. Run on a Compute Canada cluster, a single $t=2\times10^{6}$
simulations took approximately 8 hours of wall time using 16 cores.
This constitutes a roughly 200 factor speedup over a traditional CPF
implementation.

\section{Simulation Results}
\label{Simulation-Results}

Having derived a sharp interface model for cells with tunable elasticity,
we present the results of our simulations. The model can be summed
as solving Equation \ref{eq:drdt total}, using the natural curvature in Equation \ref{eq:Knatural}
and with the parameters shown in table \ref{tab:Simulations-Parameters.-Table}.
As in CPF, the model can be used at any degree of confluence, that is at any concentration, 
\[\rho=\frac{N_{cells}\pi R_{0}^{2}}{L^{2}},\]
where $L$ is the length of the simulation box. This does not account for the gap between neighboring cells that is proportional to $\lambda$ thus it slightly under-represents the actual degree of confluence.
In section \ref{subsec:Burst-Behavior}
we demonstrate our model can qualitatively recover features seen in
experiment and in the CPF model, including increased velocity bursts,
and a higher diffusion for soft-in-normal as compared to the all-normal
cells. Thanks to the large computational speedup of this model over
CPF, section \ref{subsec:Concentrations} examines the behavior across
a broad range of concentrations, varying cell stiffness, and properties
of the active motor. Notably, we show a single parameter characterization
of the instantaneous velocity distribution as a function of concentration.

\subsection{Burst Behavior\label{subsec:Burst-Behavior}}

\begin{figure}[tb]
\includegraphics[width=0.9\columnwidth]{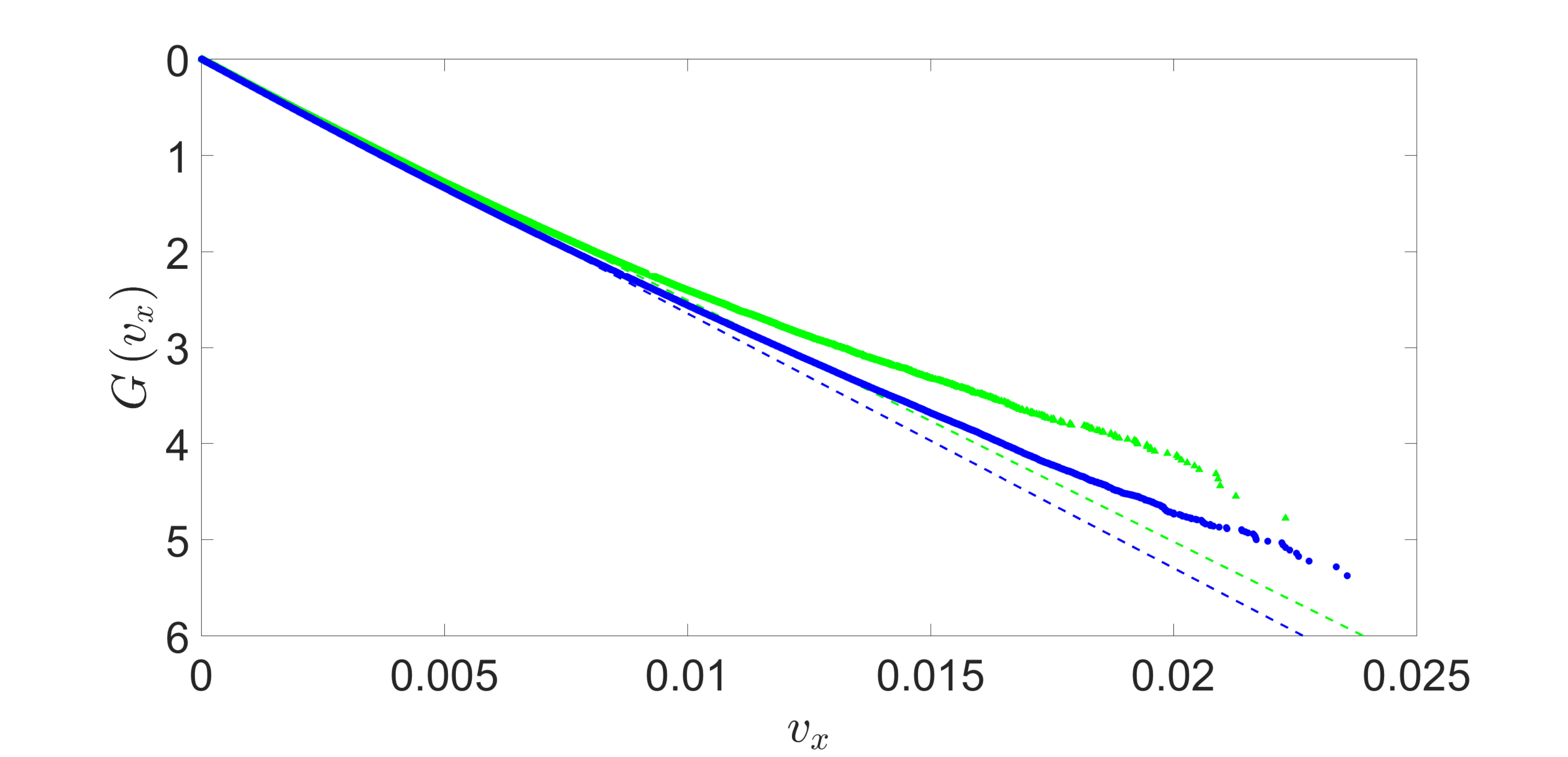}

\caption{\textbf{Sharp Interface Model Recovers Soft Cell Burst.} Probability
plot for instantaneous cell velocity of both soft-in-normal (green
triangles) and all other normal cells (blue circles) of a $\rho=83\%$
simulation. Plotted as Quantile where Gaussian Probability is a straight
line, and showing best Gaussian fit for soft-in-normal (green dashed line)
and all other normal cells (blue dashed line). \label{fig:Sharp-Interface-Model}}
\end{figure}

Rare bursts of high cell velocity were observed in experiment \cite{Lee2012}
as well as by our previous CPF model and play a key role in higher
cell motility. Thus, we begin evaluating the sharp interface model
by showing it also recovers bursts. The system consists of a single
soft cell surrounded by stiff (normal) cells, where all other cell
parameters are identical, except for the random motor orientation.
The velocity distribution of a single simulation performed at $\rho=83\%$ is shown
in Figure \ref{fig:Sharp-Interface-Model}. The probability
distribution of the absolute value of the $x $ component (or equivalently, the $y$ component) of cell velocity is plotted
as a half-normal distributed quantile, such that a half-normal distribution
would be a straight line. This is done to better show differences
in the tail of the distribution. We combined the results from $20$ independent runs to reduce noise and better illustrate differences deep in the tail of the distributions. 
The plot shows that the soft cell (green triangles) has a fatter
tail than the normal cells (blue circles). This demonstrates that
the elasticity mismatch enhances cell velocity and is consistent
with CPF results. While distribution for the normal cells shows high velocities are less probable than in the soft
cell, it is clear that this distribution also has a fatter tail compared with a normal distribution as shown by the deviation from the line of best fit. In CPF, the comparison was
performed with a new simulation replacing the soft cell with a normal
cell. However, we have found that statistically a new simulation
behaves indistinguishably from any of the normal cells in the soft-in-normal
simulation. As such, we collected data from all normal cells in the
soft-in-normal simulations, thereby reducing noise. The non-Gaussian
distribution tail of the normal cells was not seen in CPF results, likely visible due to the reduced statistics as compared to this study.

\begin{figure}[tb]
\includegraphics[width=0.9\columnwidth]{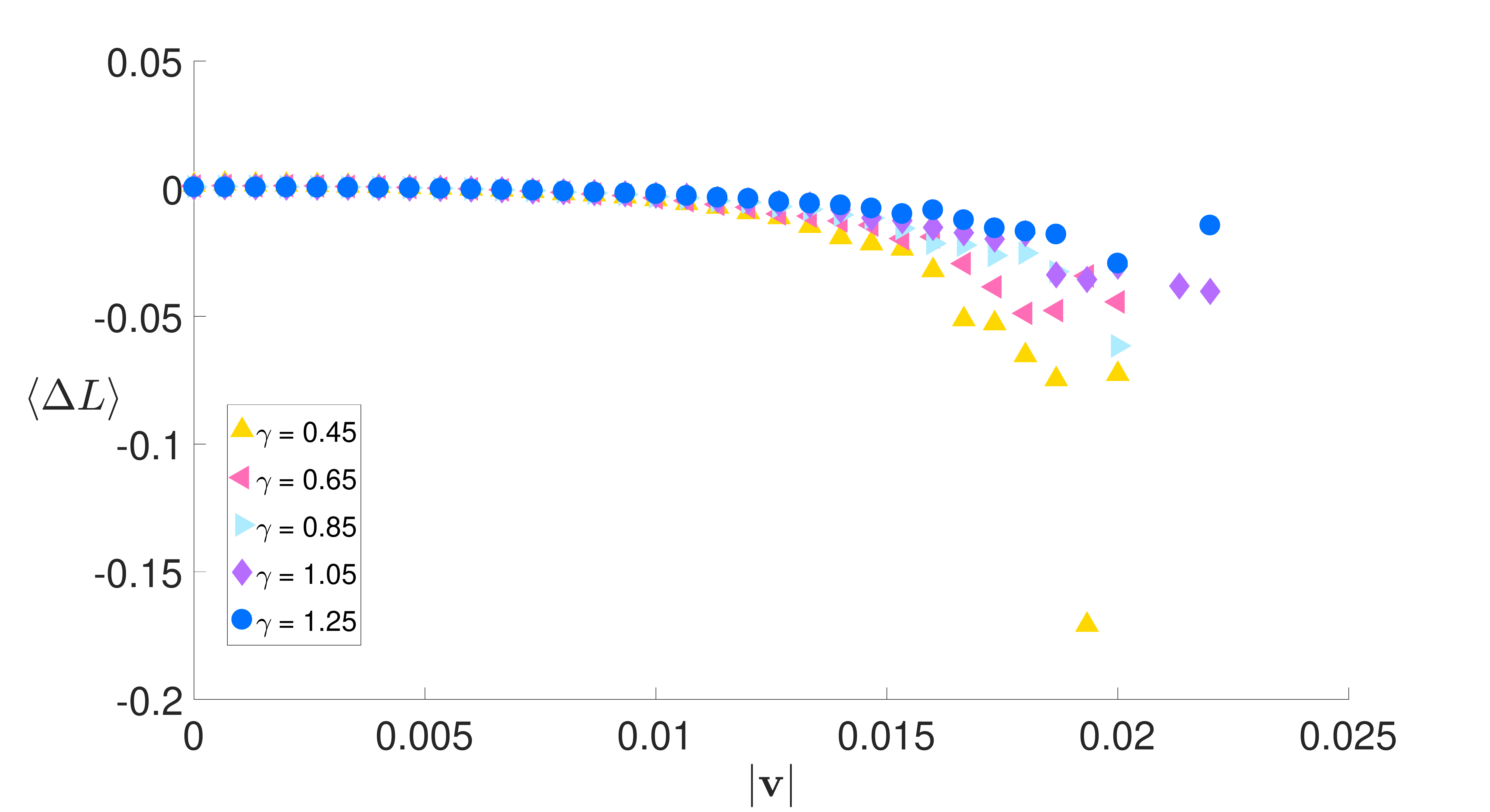}

\caption{\textbf{Average change in perimeter as a function of velocity for
different elasticities. }The average change in perimeter, $\left\langle \Delta L\right\rangle $
binned as a function of velocity magnitude $\left|\mathbf{v}\right|$. 
Each symbol corresponds to different elasticity where all
cells are adjusted, spanning $\gamma=0.45-1.25$, and the active motor speed is $0.1$. As elasticity is
decreased, higher velocities become increasingly correlated with a
negative $\left\langle \Delta L\right\rangle $, indicating that bursts
are likely to occur when the cell has contracted from a more deformed
to a less deformed state.\label{fig:Scaling-of-velocity}}
\end{figure}

To better illustrate the relation between elasticity and bursts, we
look at the average change in perimeter $\left\langle \Delta L\right\rangle $ (divided by $2\pi R_0$)
from the previous time-step, binned for different velocity magnitudes
$\left|\mathbf{v}\right|$, as shown in Figure \ref{fig:Scaling-of-velocity}.
The elasticity of all cells is changed, ranging from the all-normal cells $\text{\ensuremath{\gamma}=1.25}$ 
to all-soft cells $\text{\ensuremath{\gamma}=0.45}$,
all performed at $\rho=85\%$. For all elasticities, low velocities
are uncorrelated with perimeter change as $\left\langle \Delta L\right\rangle \approx$0.
However, higher velocities are correlated with a decrease in $\left\langle \Delta L\right\rangle $.
This indicates that bursts are more likely to occur when the cell
has relaxed from a more deformed configuration to one with a smaller
perimeter length. Comparing different series, it is also evident that
reducing cell elasticity increases this effect. As we will show later,
this increase in bursts results in a higher diffusivity for softer
cells.

We have shown that our sharp interface model can recover the main
features of the CPF model, at a significant reduction in computation
time. This speedup makes a broader study of parameter space feasible.
We begin by studying the effect of system concentration. As before,
a single soft cell with all other cells having normal elasticity are
simulated at various concentrations, ranging from near confluent $\rho=95\%$
to the very dilute $\rho=25\%$.

\begin{figure}[tb]
\textbf{(a)}\includegraphics[width=0.45\columnwidth]{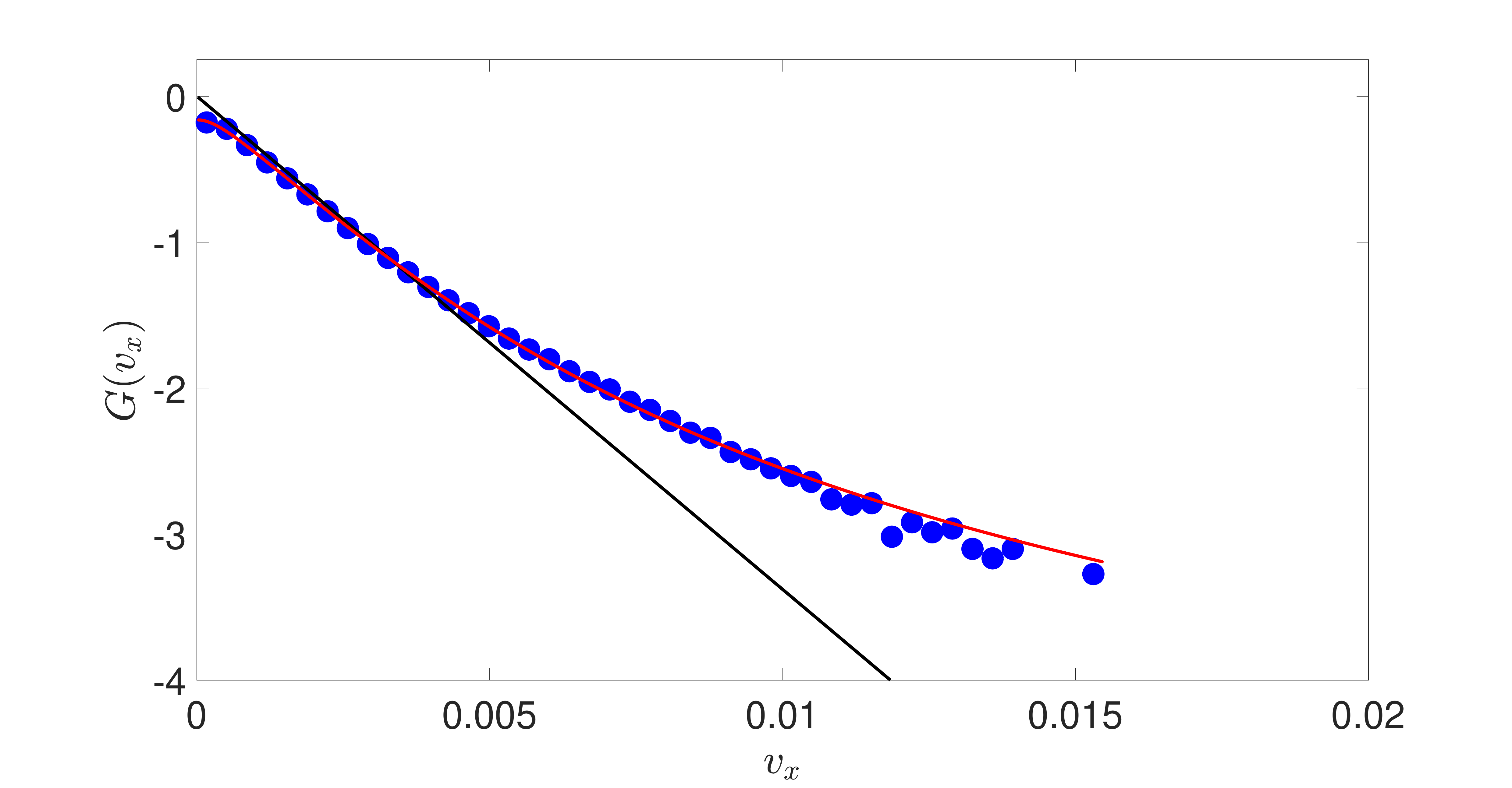}\textbf{(b)}\includegraphics[width=0.45\columnwidth]{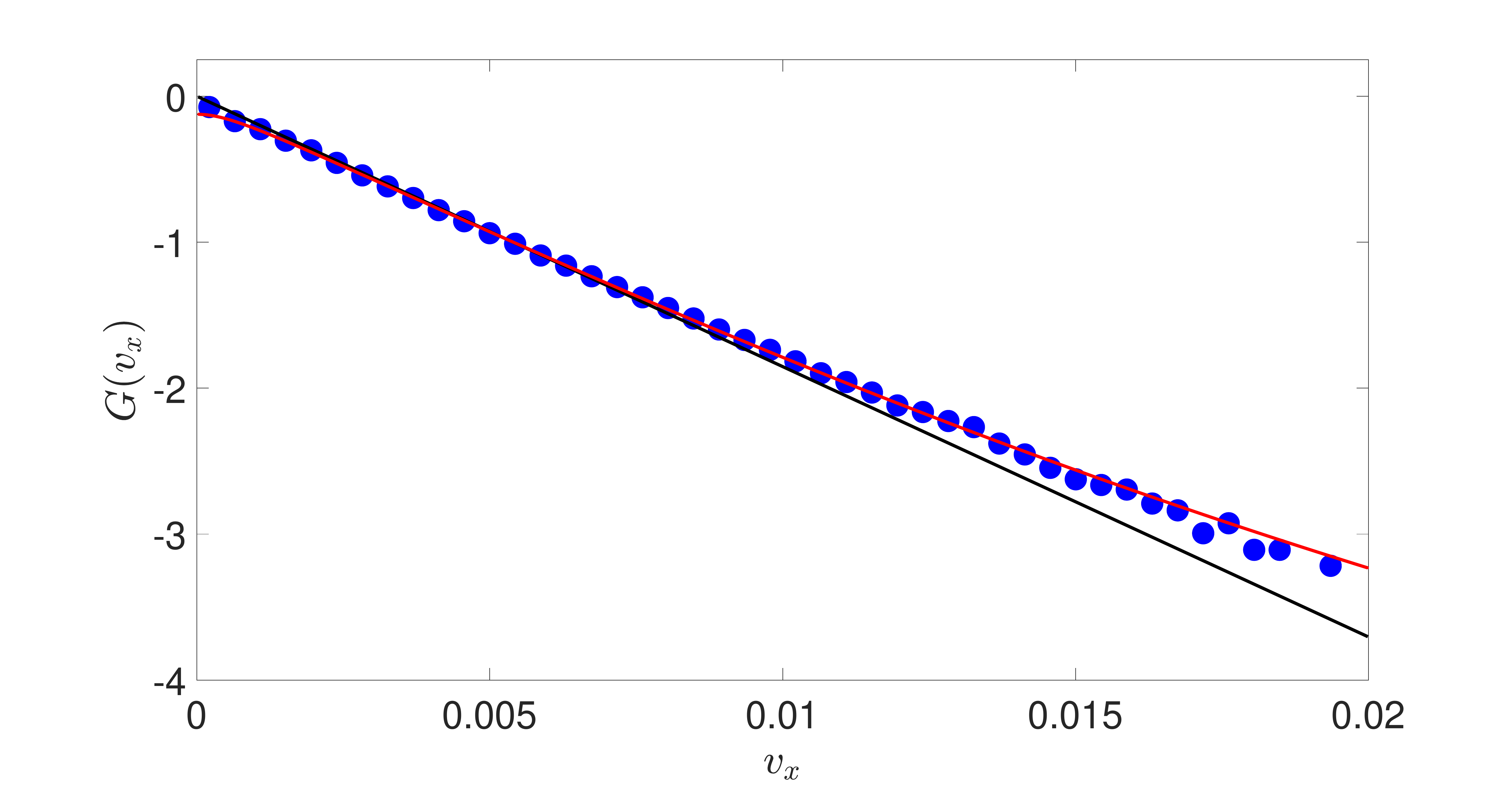}
\textbf{(c)}\includegraphics[width=0.45\columnwidth]{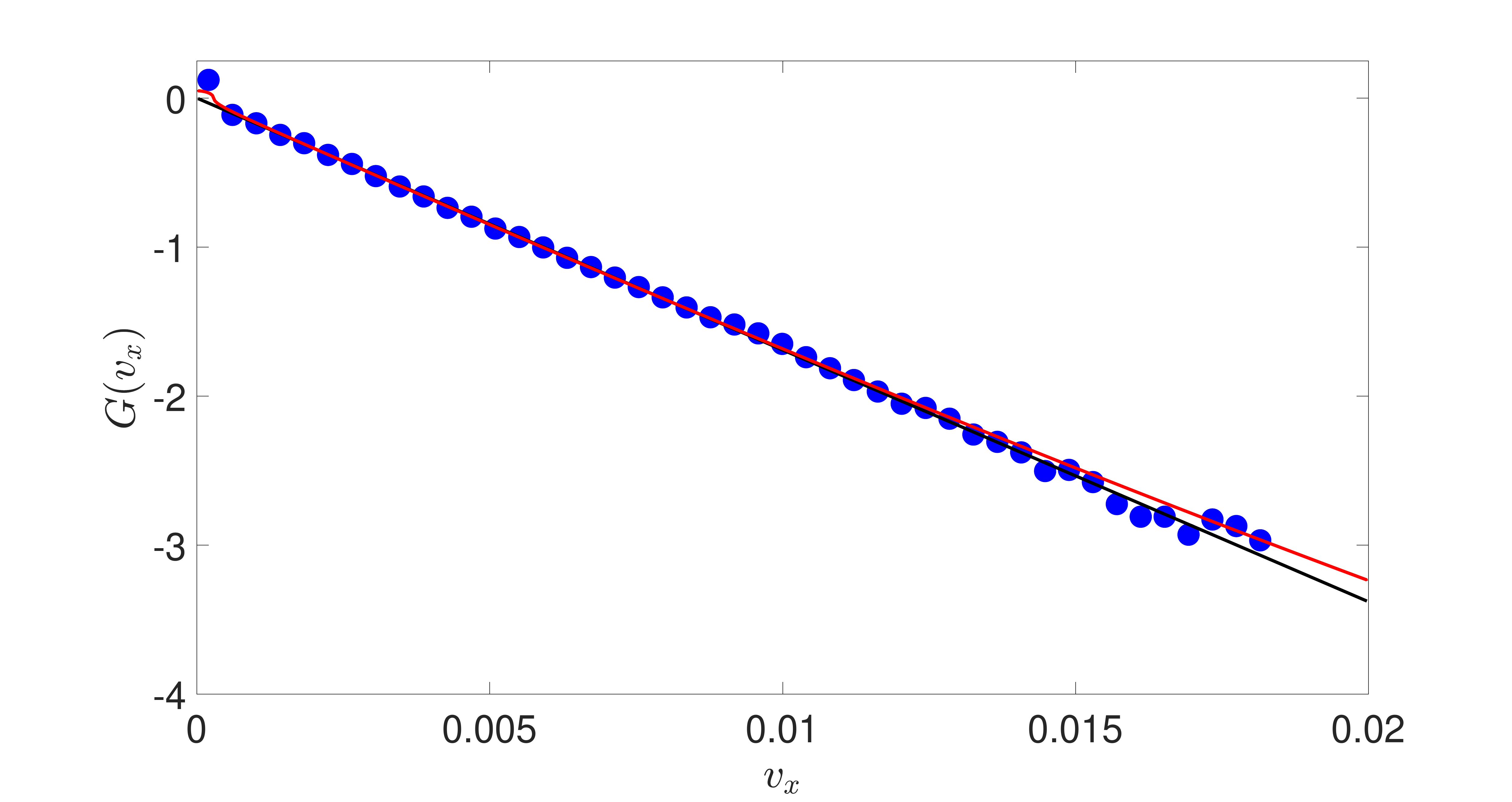}\textbf{(d)}\includegraphics[width=0.45\columnwidth]{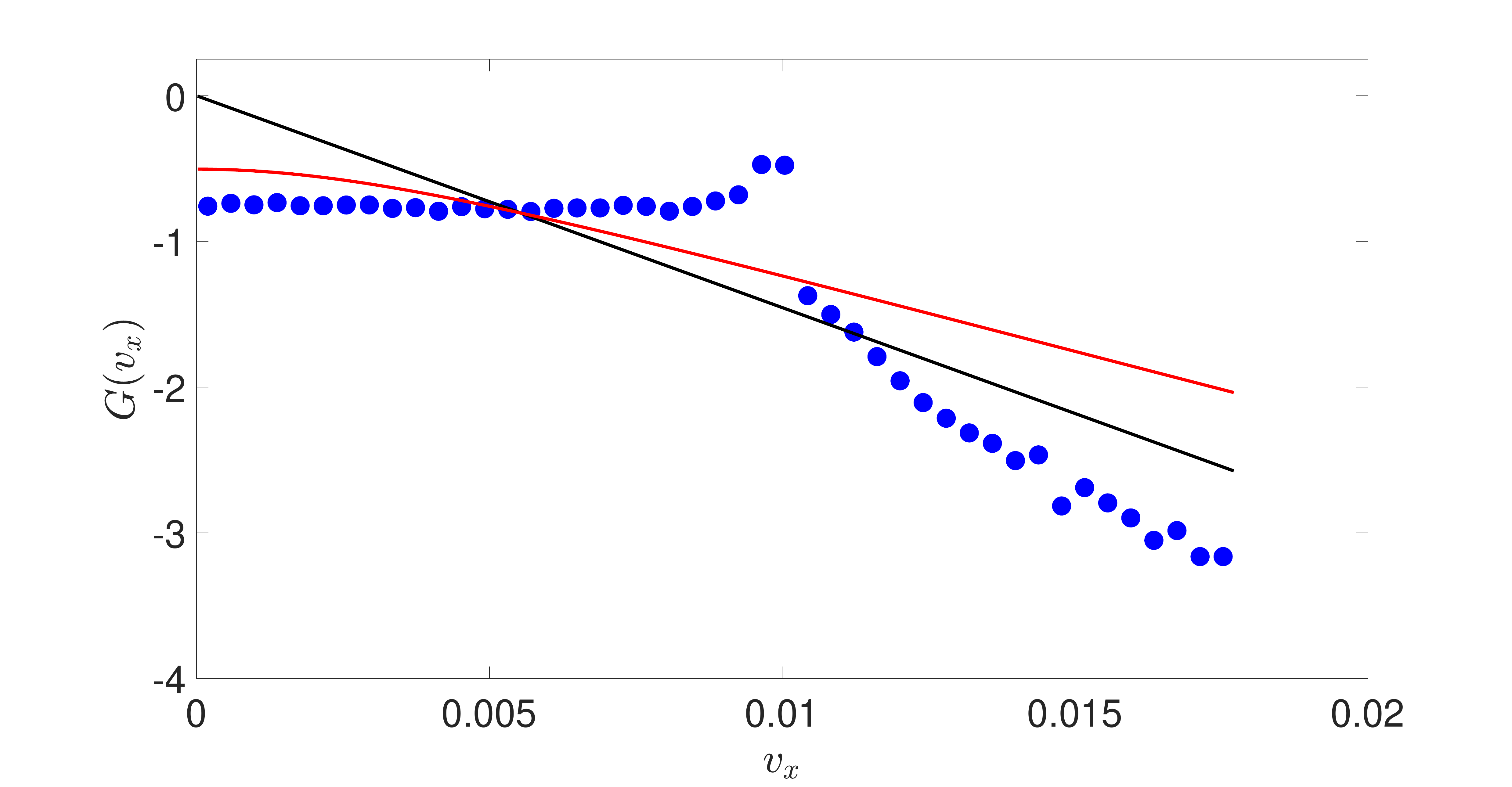}

\caption{\textbf{Velocity distribution. }The instantaneous velocity probability
distribution for all-normal cells at different concentrations (\textbf{a:} $\rho=93\%$,\textbf{
b:} $\rho=81\%$,\textbf{ c:} $\rho=77\%$, \textbf{ d:} $\rho=25\%$) plotted such that a Gaussian
distribution is a straight line. Also plotted are Gaussian (black
line) and student-t (red line) best fit lines. \label{fig:3_velocity_distributions}}
\end{figure}

Figure \ref{fig:3_velocity_distributions} shows the velocity probability
distribution of an all-normal simulation in three regimes: \ref{fig:3_velocity_distributions}\textbf{(a)
}a very dense system, $\rho=93\%$\textbf{ (}Movie\textbf{ }\#1),
where the velocity probability distribution is highly non-Gaussian,
but overall velocity is limited by the presence of neighboring cells;
\ref{fig:3_velocity_distributions}\textbf{(b \& c)} intermediate concentrations,
$\rho=81\,\&\,77\%$ \textbf{(}Movies\textbf{ }\#2 \& \#3), cells move faster but
there are less bursts as the tail is nearly Gaussian; and finally
\ref{fig:3_velocity_distributions}\textbf{(d)} at a very dilute concentration,
$\rho=25\%$ \textbf{(}Movie\textbf{ }\#4), where the cells interact less frequently and the distribution resembles that of an isolated cell, which peaks at the motor active velocity.
Each plot also shows several best fit lines. The solid lines are a
Gaussian fit, obtained by including velocities $0\leq v\leq v'$,
where the cut-off $v'$ is chosen to minimizes the Chi squared of
the fit. This was done since it is evident that the tail of the distribution
is non-Gaussian and including it in the fit skews the low velocity
behavior that does appear to follow Gaussian statistics. The dashed
red line is a student-t distribution fit for the entire velocity distribution.

\begin{figure}[tb]
\includegraphics[width=0.9\columnwidth]{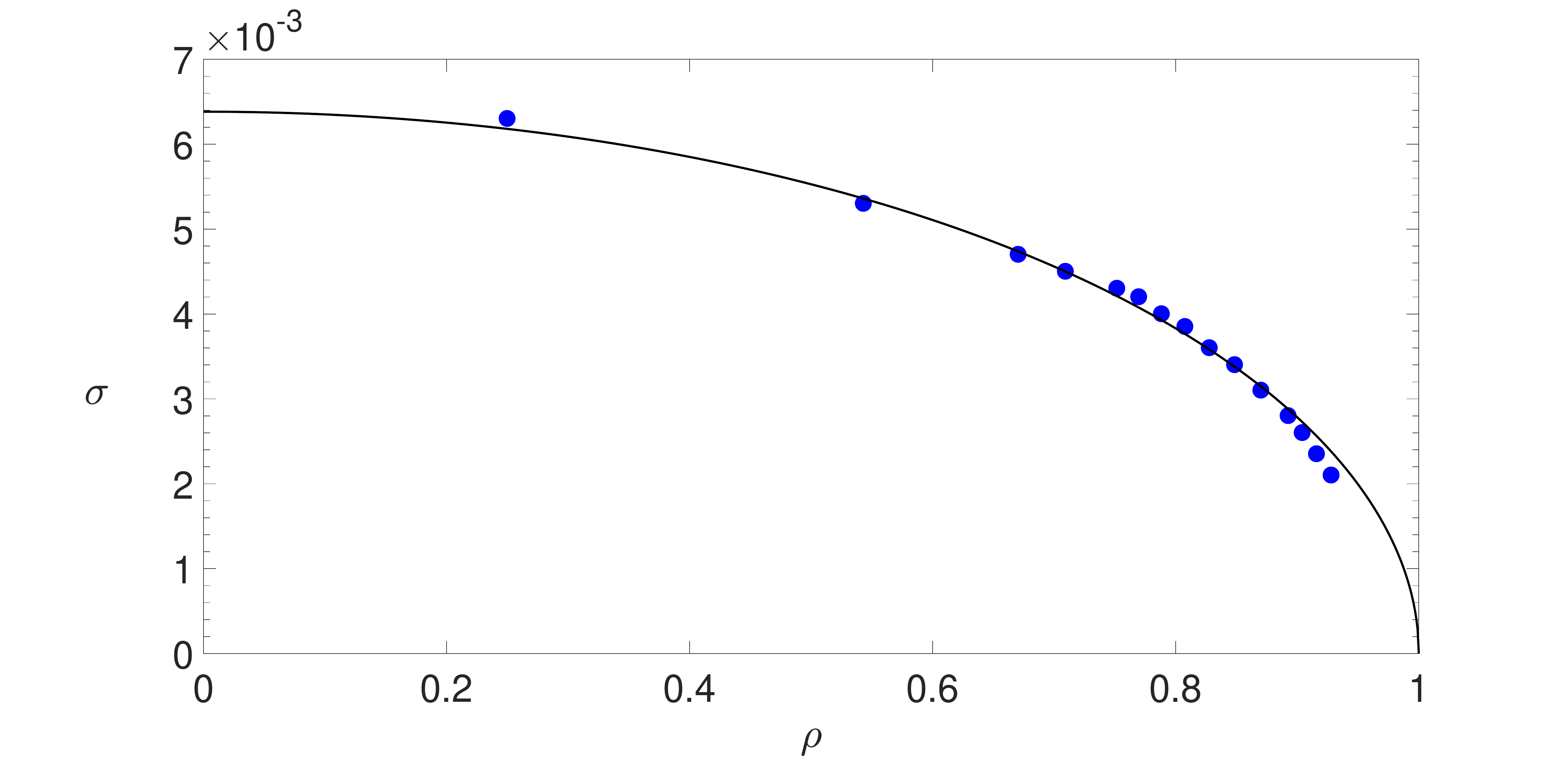}

\includegraphics[width=0.9\columnwidth]{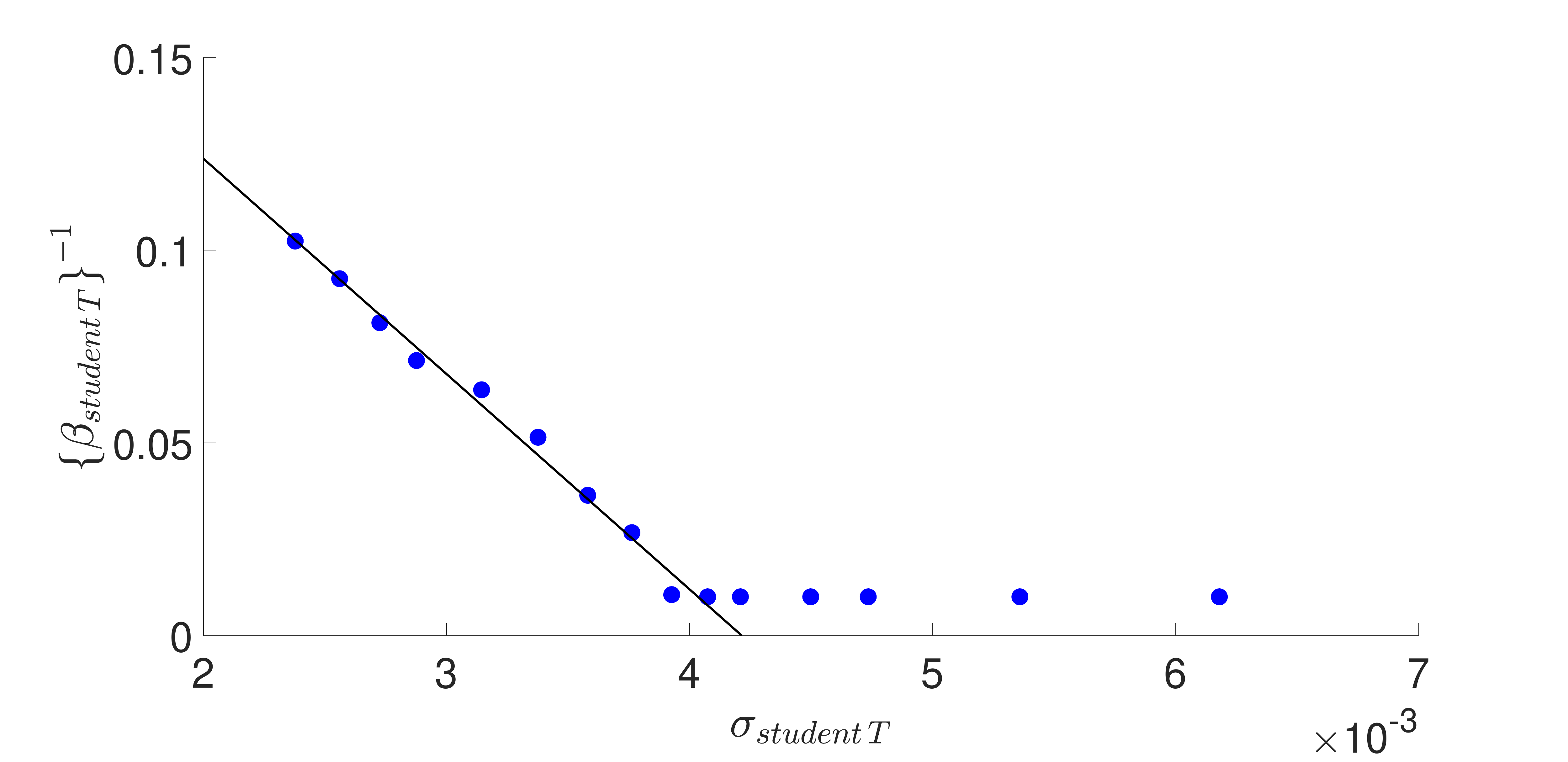}

\caption{\textbf{Student-t parameter fits across concentrations. }Student-t
fit parameter of instantaneous velocity probability distribution,
for all-normal cells at different concentrations. Top: $\sigma$ as a function of concentration,
with solid line a single parameter best fit $\sigma=a\sqrt{1-\rho^{2}}$.
Bottom: Student-t fit parameter \textbf{$\beta$} as a function of
$\sigma$ according to fit above (Equation \ref{eq:sigma by rho}). The solid line is a linear fit for small $\sigma$.\label{fig:Student-t-parameter-fits}}
\end{figure}

The student-t distribution proved to be a fit for the velocity distribution
in all but the most dilute of simulations. Given a set of Gaussian
random numbers $X_{i}$ with mean $0$ and standard deviation $\sigma$,
sampling $\nicefrac{\sum_{i=1}^{n}X_{i}}{S\sqrt{n}}$ (where
$S$ is the sample variance) gives a student-t distribution with $n-1$
degrees of freedom. $X_{i}$ can be thought of as the cell motor sampled
over the neighboring cells that it is interacting with. Here we use
the 2-parameter student-t, where $\sigma$ sets the scale of the distribution,
and $\beta=n-1$ denotes the degrees of freedom \cite{jackman2009bayesian}. Results of fitting
the student-t $\sigma$ parameter are plotted as a function of concentration
at the top of Figure \ref{fig:Student-t-parameter-fits}. This shows
a clear relation where increased $\rho$ leads to lower $\sigma$,
as crowding makes higher velocities increasingly unlikely. We found
our results are well described by the one parameter fit 
\begin{equation}
	\sigma\left(\rho\right) = a\sqrt{1-\rho^{2}},
	\label{eq:sigma by rho}
\end{equation}
shown as a solid line. Therefore, we motivate this particular form
by considering the symmetry around $\rho=1$. Although as plotted
here $\sigma$ is positive semi-definite, the complete $v_{x}$ distribution
is both positive and negative, and symmetric around zero. Similar
to a Gaussian distribution, $P\left(v=\pm\sigma\right)/P\left(v=0\right)\approx1/\sqrt{e}$
(Given the relation for $\sigma\left(\rho\right)\&\beta\left(\rho\right)$,
this is within 5\% even at $\rho=1$), and this form preserves this
symmetry continuity. In the bottom of Figure \ref{fig:Student-t-parameter-fits},
we plot the other fit-parameter of the student-t, $\beta$. Rather
than as a function of concentration, $1/\beta$ as a function of the expected $\sigma$ from Equation \ref{eq:sigma by rho}
shows that below a critical value of $\sigma*$, it follows
\begin{equation}
\left.\beta\left(\sigma\right)\right|_{\sigma < \sigma*} = \frac{1}{b_1 \sigma + b_0}.
\label{eq:beta by sigma}
\end{equation}
This reduces the two parameter
student-t to a single fit parameter $\sigma$ (or equivalently, $\rho$).
As concentration is increased, cells are more confined and move with
reduced velocity, yet the tail of the distribution becomes more pronounced
as collective behavior leads to increased bursts. Note that fitting beyond $\beta=100$ proved difficult
as it is numerically indistinguishable from a Gaussian distribution.
There are, of course, many other possible distributions that resemble
a Gaussian with fatter tail. In our previous analysis of the CPF results,
we had shown that the heavier than Gaussian tail of the velocity distribution
could be fitted in two ways: the student-t distribution, as well as
a single parameter 'caged cell' fit where the soft cell mostly behaves like a normal cell, with the exception of rare events where
its motor supplements the Gaussian velocity given by the surrounding
cells. Here we found the caged cell model was not a good fit, particularly
for the all-normal cells. 

\begin{figure}[tb]
\includegraphics[width=0.9\columnwidth]{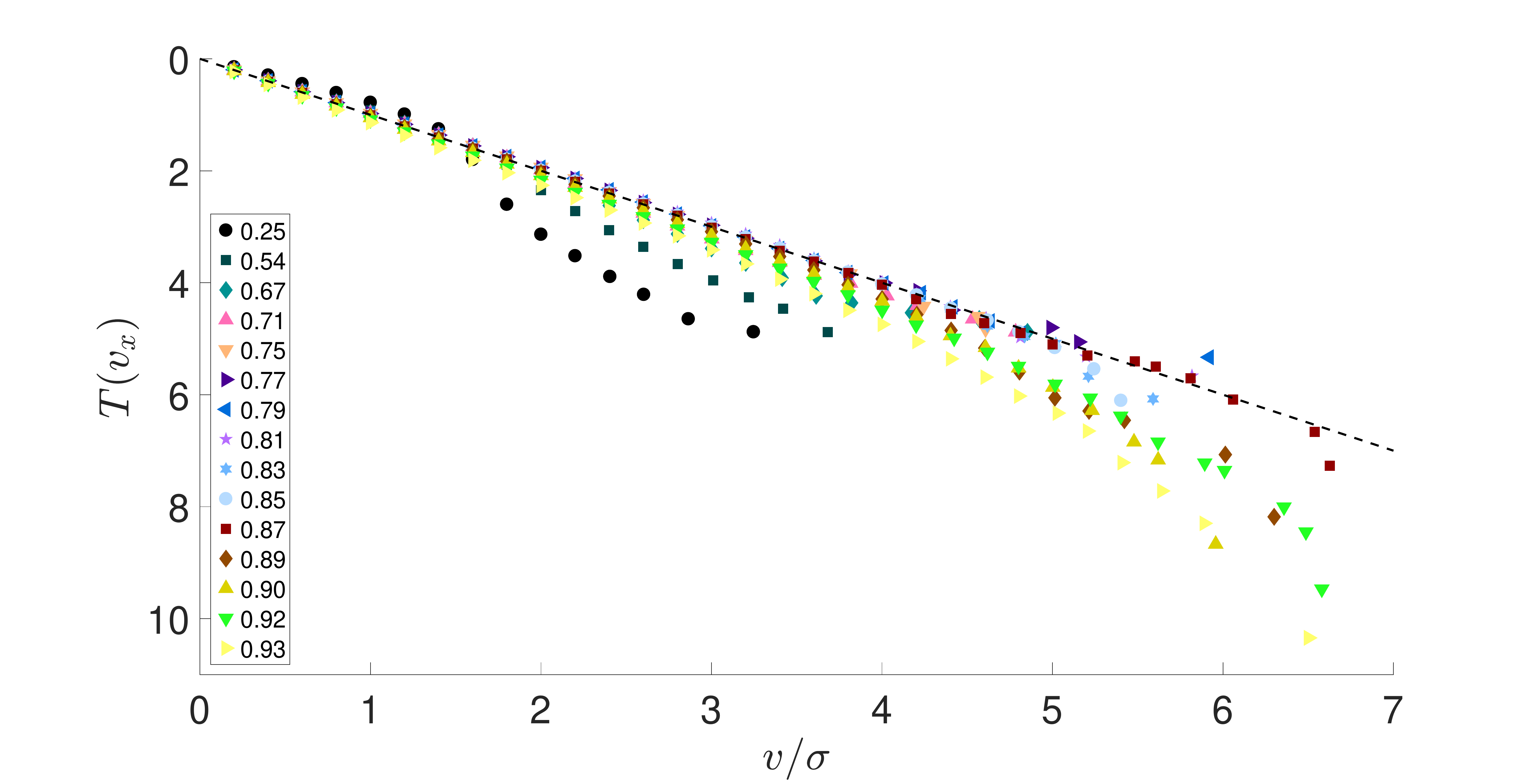}

\caption{\textbf{Instantaneous velocity probability distributions master curve.} Velocity profiles for all-normal cells at different concentrations (shown in the legend) rescaled such that a student-t distribution forms a straight line. Each concentration was scaled by $\sigma\, \&\, \beta$ given by Equations \ref{eq:sigma by rho} $\&$ \ref{eq:beta by sigma}. The dashed line represents an ideal student-t distribution (for any $\sigma\, \&\, \beta$). Data collapses onto the dashed line, with the exception of the $2$ most dilute simulation.  \label{fig:Velocity-distribution}}
\end{figure}

To further illustrate the scaling of the velocity distribution with
respect to concentration, Figure \ref{fig:Velocity-distribution}
shows the velocity probability distribution for all concentration in a quantile plot $T\left(v_x\right)$ 
such that a student-t distribution would form a straight line. 
Each concentration was rescaled by $\sigma\left(\rho\right), \beta\left(\sigma\right)$ as given 
from Equations \ref{eq:sigma by rho},\ref{eq:beta by sigma} respectively.
The data largely collapses to the master curve showing that the student-t parameter 
relations as a function of $\rho$, hence these are accurate in predicting the full velocity distribution at any concentration.
The slight spread at deep in the tail of the distribution is likely due to the small sampling of rare events, and propagation of error in determining $\sigma\,\&\,\beta$. 
The main exceptions are the two most dilute simulations. 
These more closely resemble the distribution of an isolated cell, with a peak at $v=v_{A}=0.01$, as was seen
in Figure \ref{fig:3_velocity_distributions}\textbf{(d)}. Here, the sharp fall off cannot be described by a normal or student-t distribution.

\subsection{Cell Motility\label{subsec:Concentrations}}

\begin{figure}[tb]
\includegraphics[width=0.9\columnwidth]{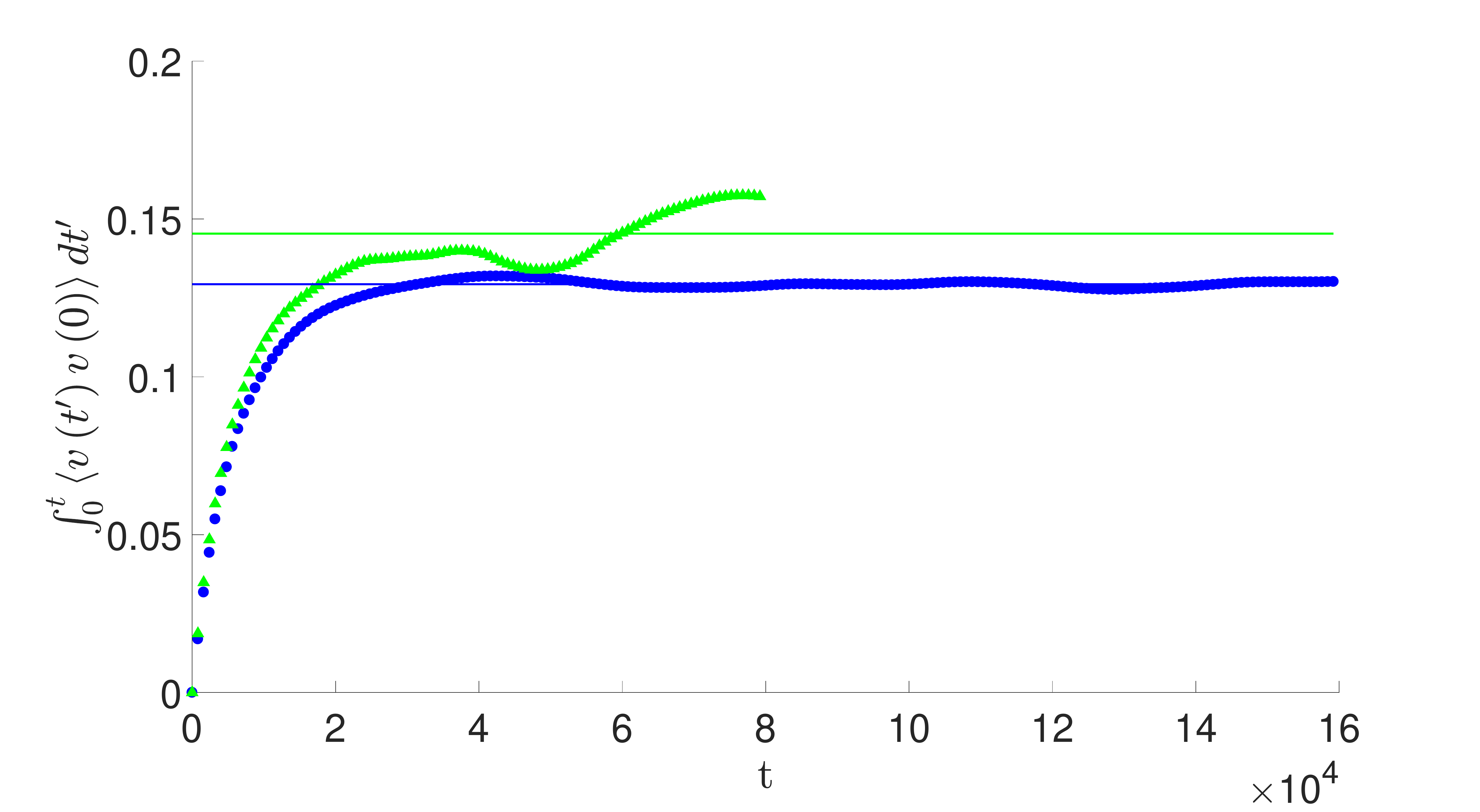}

\caption{\textbf{Estimating Diffusion constant from Velocity auto-correlation.
}Integrating the velocity auto-correlation function (VACF) of soft-in-normal
(green triangles) and all other normal cells (blue circles) is used
to estimate the diffusion constant, $\rho = 87\%$. Averaging over later times is
used to obtain an estimate for the diffusion constant, shown as a
solid green (blue) line for soft-in-normal (normal) cells. \label{fig:estimatingD}}
\end{figure}

Having shown that softer cells experience more bursts, we now examine
whether this leads  to higher cell motility. There are various
methods for calculating the motility. Here we chose to use the integral
of the velocity auto-correlation (VACF), $D\left(t\right)=\frac{1}{2}\int_{0}^{t}\mathbf{\left\langle v\left(\mathrm{t}\mathrm{'}\right)\cdot v\left(\mathrm{0}\right)\right\rangle \mathrm{dt'}}$,
where taking $\lim_{t\rightarrow\infty}D(t)$ yields the diffusion
constant \cite{Green1954,Kubo1957}. We estimate this by averaging over
$D\left(t\right)$ at late times. As shown in Figure \ref{fig:estimatingD},
this allows for a good estimate as the VACF quickly converges to $0$,
and so a shorter time window can be used to accurately estimate $D\left(\infty\right)$
by averaging $D\left(t\right)$ after it stops rising. The solid line
shows the final result of averaging. Since these simulations include
$71$ normal cells but only $1$ soft cell, the latter has significantly
more noise. The fluctuations of the average underestimate the systematic
error since they are correlated. Instead we estimated systematic error
as follows: performing $20$ independent simulations, a diffusion
constant was computed for each run. Fitting a normal distribution
to these 20 results yields their standard deviation. We find that
it is $3.2\%$ for the normal cells, and $9.2\%$ for the soft-in-normal
cell. As such, it is difficult to evaluate the behavior of the
soft-in-normal cell, beyond what we already demonstrated in Figure \ref{fig:Sharp-Interface-Model}.

\begin{figure}[tb]
\textbf{(a)}\includegraphics[width=0.9\columnwidth]{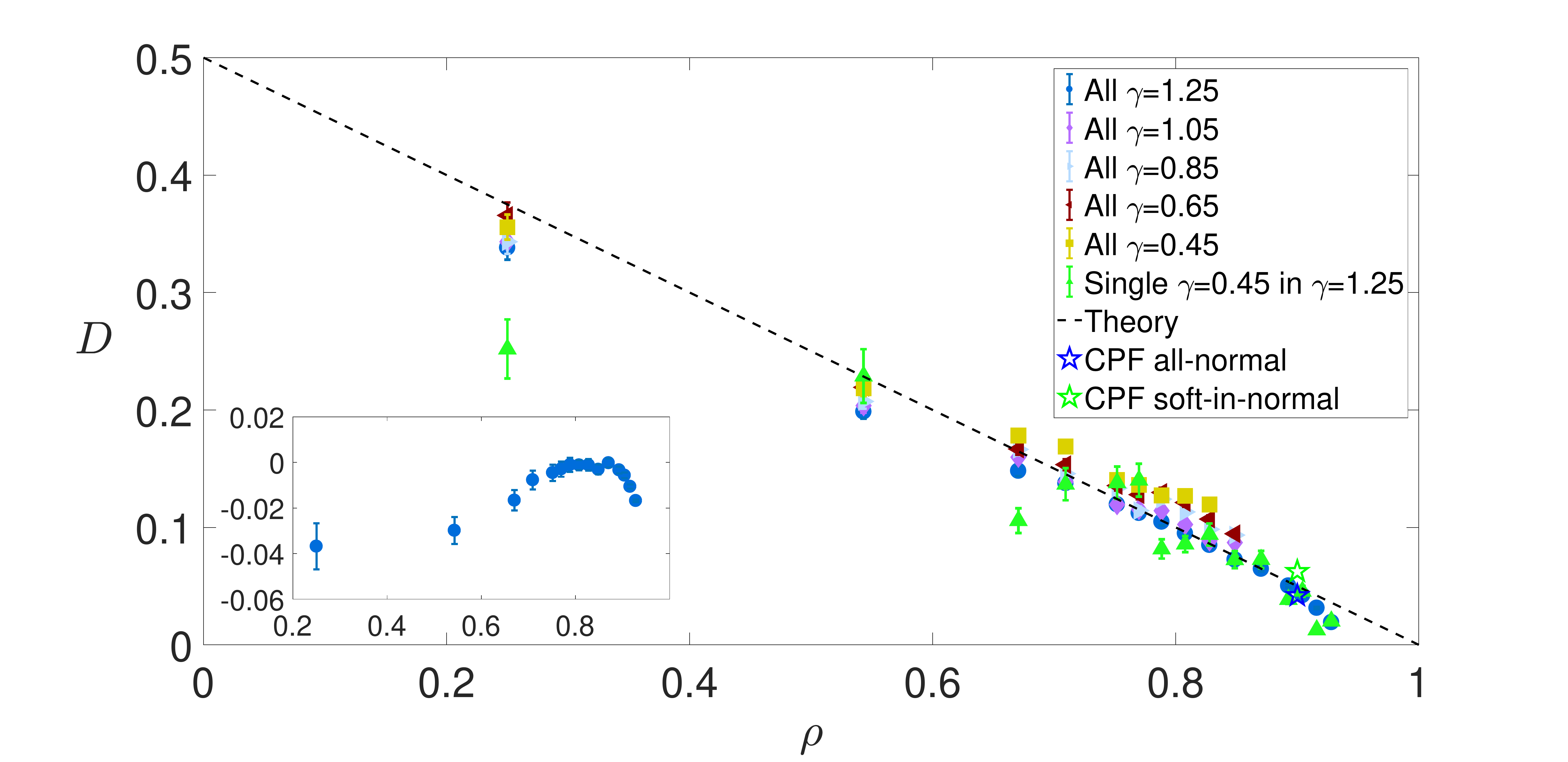}

\textbf{(b)}\includegraphics[width=0.9\columnwidth]{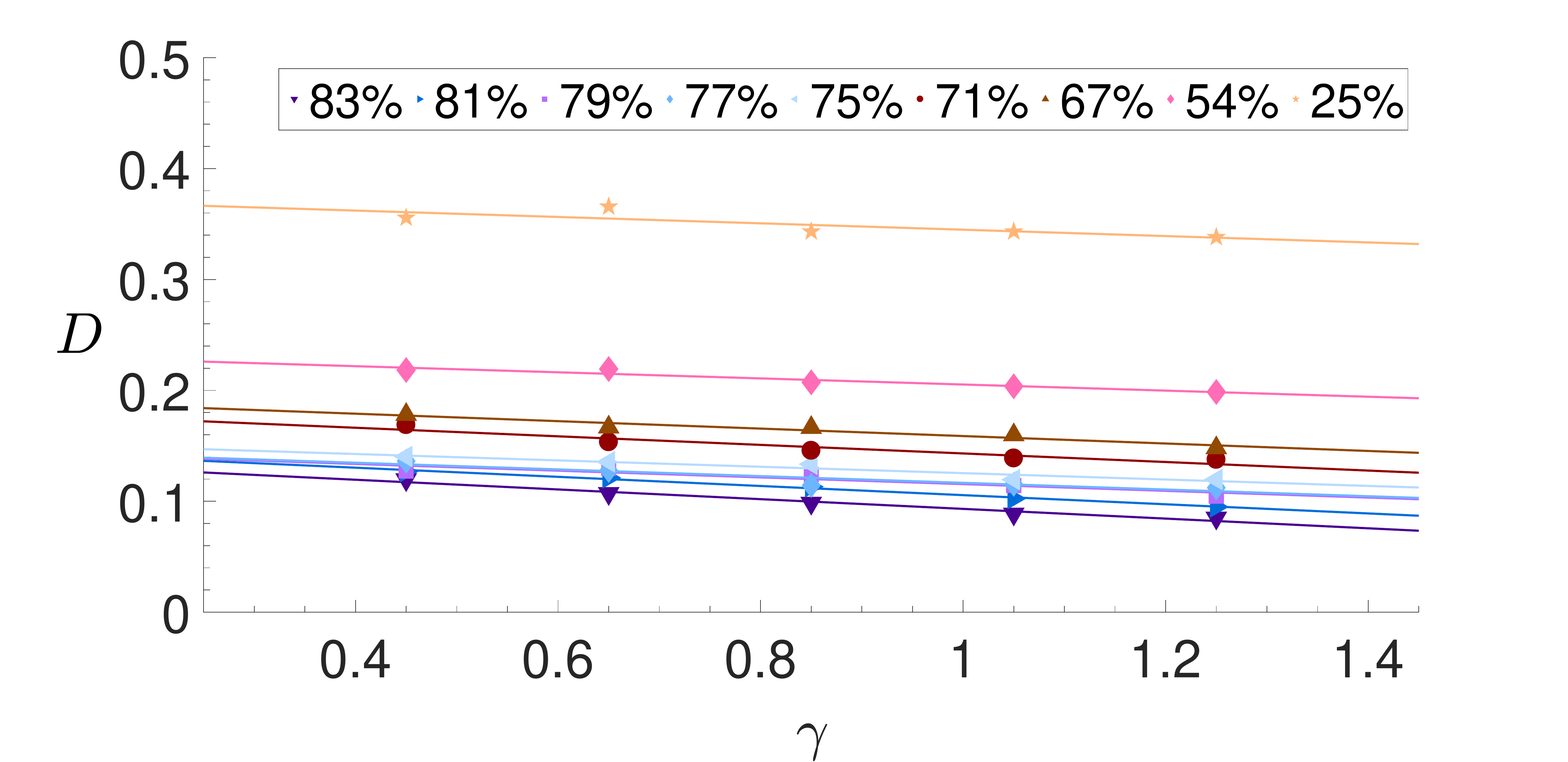}

\caption{\textbf{Cell diffusion as a function of concentration and elasticity.}
\textbf{(a)} Diffusion constant as a function of concentration, and
for various elasticities, as shown in the legend. Also shown is the well known theoretical result Equation \ref{eq:DrhoTheory}, as well as results from
CPF for both soft (green star) and normal (blue star). \textbf{(b)}
Diffusion constant as a function of elasticity, at different concentration as shown in legend,
with linear best fit line for each series. \textbf{\label{fig:Diffusion-constant-from}}}
\end{figure}

Examining the instantaneous velocity distribution, we showed competition
between increasing burst frequency and an overall reduction in the
mean velocity as $\rho$ increases. One may ask how does this effect cell motility?
Following the procedure outlined previously, Figure \ref{fig:Diffusion-constant-from}\textbf{(a)}
shows the motility for different concentrations, as well as for different
cell stiffness. Compared to the all-normal case, error for the soft-in-normal
is considerably higher, which makes it difficult to draw conclusions.
Therefore, to study the effect of elasticity, it was varied for all cells in
a given simulation ranging from the all-normal $\gamma=1.25$ to the
all-soft $\gamma=0.45$. Also shown on the plot are our two previous
CPF results, for both the soft-in-normal (green star) and normal-in-normal
(blue star) simulations at $\rho=90\%$, and the well-known result 
\begin{equation}
D(\rho)=D_{0}(1-\rho),
\label{eq:DrhoTheory}
\end{equation}
 derived for a discrete random walk where cells move to an adjacent
site unless it is already occupied \cite{phillips2013}. Overall, our results are
consistent with these three results. The linear fit
appears to describe general scaling, and diffusion is reduced with
increased concentration. The inset shows the residuals between the
$\gamma=1.25$ all-normal simulation and the theoretical fit. Though
the residuals are small, there appears to be some additional higher
order behavior beyond linear scaling. We also see that at each concentration,
lower elasticity leads to increased diffusion. To better illustrate
this, in Figure \ref{fig:Diffusion-constant-from}\textbf{(b) }we
plot $D\left(\gamma\right)$ for different concentrations. The best
fit lines correspond well with data, indicating that at all concentrations, 
increased elasticity reduces cell diffusion, at a rate that is roughly
constant within error. Hence, decreasing cell elasticity leads to increased
cell motility, at all concentrations, consistent with results from
CPF.

\begin{figure}[tb]
\includegraphics[width=0.9\columnwidth]{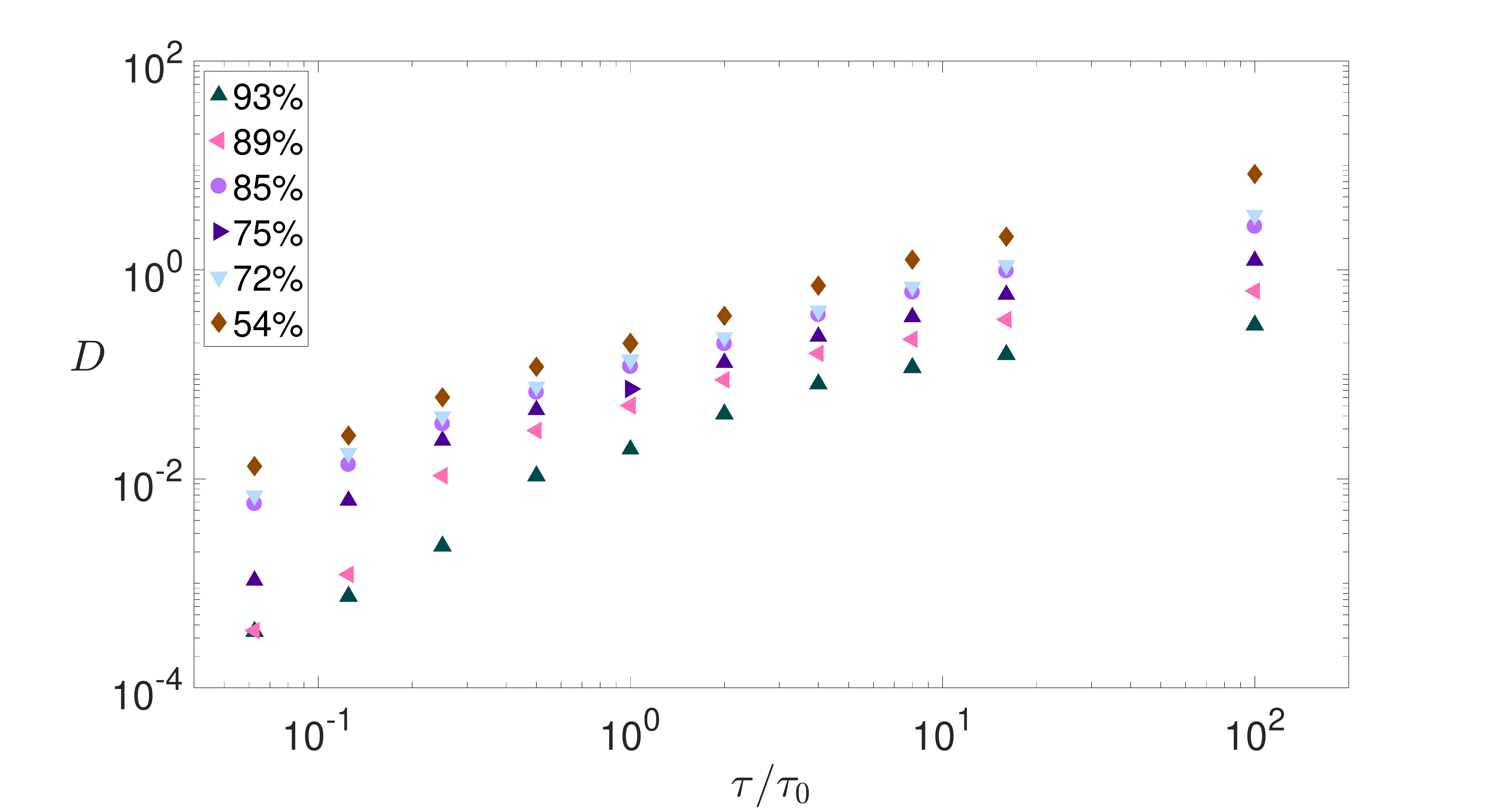}\caption{\textbf{Cell diffusion as a function of motor reorientation time $\tau$ and
concentration.} Cell motility increases with increased $\tau/\tau_0$, but in a non-linear fashion, for different concentration, as shown in legend. \label{fig:DiffusionAcrossTauAndConcentration} }
\end{figure}

Finally, we examine the role of the cellular motor parameters 
on diffusion. For isolated cells, diffusion can be derived analytically, $D_{iso}=\frac{1}{2}v_{A}^{2}\tau$.
For non-isolated cells, that relation no longer holds for several
reasons. As cells are interacting, the effective velocity is on average
lower than the active velocity. As well, reorientation time due to
collisions may be shorter and dominate over the motor reorientation
time. To decouple these effects, Figure \ref{fig:DiffusionAcrossTauAndConcentration}
shows the resulting diffusion constant when $\tau$ is varied while
keeping $v_{A}$ constant, for a range of concentrations. $\tau_{0}$
is the reorientation time used previously in this paper, as shown
in table \ref{tab:Simulations-Parameters.-Table}. For $\tau/\tau_{0}\leq1$,
the diffusion constant was calculated as in previous sections, by
averaging over the tail of $D(t)=\int_{0}^{t}\left\langle v\left(t'\right)v\left(0\right)\right\rangle dt'$.
However, for $\tau/\tau_{0}>1$ the diffusion constant is large, and
this method is slow to converge. Instead, an exponential fit to $D\left(t\right)$
is performed to extrapolate the long-time diffusion constant. As seen
in the plot, though the particular value of $\tau$ affects the magnitude
of cellular diffusion, the qualitative behavior of higher $\rho$
leading to slower diffusion remains consistent. It is also evident
that the measured diffusion is non linear with respect to $\tau$,
and the slope at $\tau=\tau_{0}$ is smaller than 1. At $\tau/\tau_{0}\ll1$,
the active velocity oscillated rapidly, leading rapid decorrelation of the VACF and hence, a decreased diffusion constant \cite{Hurley1995}. Due to collisions with the
surrounding cells, increasing $\tau$ increases $D$ but at a diminishing
rate. This reduction is correlated with increased concentration, and
diffusion appears to level off at a lower value of $\tau/\tau_{0}$
for denser systems.

\section{Discussion}

Our paper focused on a new, sharp interface model for the simulation
of assemblies of motile cells with varying elasticities. The model was inspired by a previous
phase field approach to this system. While successful in recreating
velocity bursts, this method was very computationally demanding. By
approximating the cell interface and ignoring all bulk behavior, we
were able to dramatically speed up simulation time by a factor of
$\sim200$. Model parameters allow to independently tune differing
elasticities for each cell while keeping all other cell properties identical.
These parameters were largely derived from the CPF model, using equilibrium
approximations to determine the strength of the various terms. The
sharp interface model is more susceptible than CPF to cell pinching,
where opposite ends of the cell interface can overlap. In this study
we circumvented this issue by making the cells slightly stiffer and
adding a term which corresponds to a spontaneous curvature in the energy functional. 
Other solutions, such as adding an internal cell wall repulsion are feasible and may allow
the simulations of softer cells. A comparison of these approaches
may elucidate some of the observed difference between this sharp interface
model and previous CPF results. Our model is also distinct from other
sharp interface models of cells. Though some studies focused on the
motion of individual cells in greater detail, we are able to simulate
large systems efficiently, while maintaining description at the individual
cell level. Conversely, other methods that do allow for much larger
systems size do not allow for large deformations of individual cells,
which we have shown to be key to velocity bursts.

We demonstrated that this model recovers behavior seen in experiment,
as well as many but not all of the features of the full CPF model
with respect to cell dynamics. As before, soft cells show an increased
likelihood of high velocity burst events than stiffer cells. These
bursts were described by performing a student-t fit to the velocity
probability distribution, where a lower degree of freedom, $\beta$
represents a ``fatter'' tail of the distribution. We have also shown
that these bursts occur as a highly deformed cell relaxes to a more
spherical configuration, qualitatively consistent with previous CPF
results as well as with experiment. These bursts result in softer
cells having higher motility (diffusion constant) than normal cells
at the same concentration. 

We also obtained several new results that were not feasible with the
computationally slower CPF model. For a given elasticity, we have
shown that both $\beta$ and $\sigma$ are consistent with a simple
relation to $\rho$, and that these reduce the student-t to
a single parameter fit. Below a cut-off concentration, $\beta$ is
infinite suggesting the tail is Gaussian or below-Gaussian, as seen
at very dilute simulations. Above this cut-off, $1/\beta$ is linearly
correlated with concentration as more bursts are seen. On the other hand, $\sigma$
decreases with increased concentration, as cells lack free space to
move to and overall mobility is reduced. This becomes clear when examining
cell motility which appears to linear decrease with $\rho$, consistent with theory. Furthermore,
this linear relation appears to hold across various cell elasticities.
Finally, we have shown that at a given concentration, motility is
also linearly related to cell elasticity. Although softer cells are
more motile than stiffer cells, the quantitative difference between
the two is less pronounced than that in the CPF model. Finally, we
showed that varying our motor parameters does not affect these relations
qualitatively.

There are several other prospects for application of this model.
The first is to better understand some of the differences between CPF and
this sharp interface model, though this will likely require considerable
computational resources to perform comparable CPF simulations at various parameters
and with sufficient statistics. Another avenue is to extend this model
to other biologically relevant systems. In particular, future work
will examine the additions of cell-cell adhesion, which can be different
for cancer and healthy cells. It remains
to be seen whether adhesion changes the burst behavior and in what
way. Similarly, different cellular motor schemes could also lead to
changes in bursts or motility.

Finally, we are also interested in applying the model to physics systems
which may not be directly tied to biology. For example, as the concentration
tends to 1 and $\sigma$ approaches zero, we are studying an apparent ordered-disordered
phase transition, similar to those studied in other model systems.

In summary, we have developed a new model for simulating cells on a monolayer. This sharp interface model has similar advantages to CPF by explicitly tracking deformations of individual cells with tunable elasticity, yet it is $200$ times faster. We recover previous results for velocity bursts, as well as demonstrate new description of the velocity distribution and cell motility as a function of concentration and cell elasticity.

\section{Acknowledgments }

The Natural Sciences and Engineering Research Council of Canada and the Fonds qu\'eb\'ecois de la recherche sur la nature
et les technologies are gratefully acknowledged for funding this research.
Computations were performed on the Guillimin supercomputer from McGill
University, managed by Calcul Qu\'ebec and Compute Canada. The operation
of this supercomputer is funded by the Canada Foundation for Innovation
(CFI), the minist\`ere de l'\'Economie, de la science et de l'innovation
du Qu\'ebec (MESI) and the Fonds de recherche du Qu\'ebec - Nature et
technologies (FRQ-NT).

\bibliography{Automatically_Imported} 

\end{document}